\documentclass{article}
\usepackage[main, final]{neurips_2025}
\usepackage[utf8]{inputenc}
\usepackage[T1]{fontenc}
\usepackage{hyperref}
\usepackage{url}
\usepackage{booktabs}
\usepackage{amsmath,amssymb}
\usepackage{nicefrac}
\usepackage{microtype}
\usepackage{xcolor}
\usepackage{graphicx}
\usepackage{wrapfig}

\title{Differentiable N-body code for Galactic Dynamics Odisseo}
\workshoptitle{Differentiable Systems and Scientific Machine Learning}

\author{
  \setlength{\tabcolsep}{2pt}
  
  \vspace{-1mm}
  Giuseppe Viterbo$^{1}$ \quad
  Tobias Buck$^{1}$\\[2pt]
  $^{1}$\footnotesize{Interdisciplinary Center for Scientific Computing, Heidelberg University, Germany}\\
  \footnotesize\texttt{giuseppe.viterbo@iwr.uni-heidelberg.de}\\
  \footnotesize\texttt{tobias.buck@iwr.uni-heidelberg.de}
  }

\begin{document}

\maketitle

\begin{abstract}
  We introduce \textsc{Odisseo} (Optimized Differentiable Integrator for Stellar Systems Evolution of Orbits), a differentiable N-body code designed to constrain the gravitational potential of the Milky Way (MW) through dynamical modeling of accreted structures such as stellar streams. \textsc{Odisseo} is implemented in JAX, enabling just-in-time compilation, automatic differentiation, and hardware acceleration on GPUs and TPUs. The code features efficient, fully vectorized force calculations and exhibits near-linear scaling when distributing a single simulation across multiple GPUs, making it suitable for large-scale optimization tasks. As a demonstration, we present a case study using a mock GD-1 stellar stream simulation, where we optimize four physical parameters via gradient descent: the accretion time and progenitor mass, as well as the masses of the host galaxy’s Navarro-Frenk-White (NFW) halo and Miyamoto–Nagai (MN) disk. \textsc{Odisseo} accurately recovers stream morphology and underlying parameters in a differentiable and scalable framework, providing a powerful tool for dynamical studies of the Milky Way and its accreted substructures. The code is available on GitHub: \url{https://github.com/vepe99/Odisseo}. 
\end{abstract}

\section{Motivation and Related Work }
\label{sec:N-body_codes}

\paragraph{Stellar streams}
In the hierarchical galaxy‐formation paradigm, large galaxies like the Milky Way are expected to host many stellar streams – long, thin tidal debris from disrupted dwarf galaxies and globular clusters \cite{bonaca_stellar_2024}. As orbiting satellites lose stars to tidal forces, they create dynamically cold stellar streams that roughly trace the progenitor’s orbit. These streams remain coherent over billions of years, making them extremely sensitive to even small perturbations by the Galactic potential or dark substructures. These features establish stellar streams as powerful probes of the Milky Way’s gravitational field and dark‐matter distribution. 

\paragraph{N-body codes}
Our motivation in developing \textsc{Odisseo} was to combine the expressivity of full N-body simulations with the differentiability of modern machine-learning–oriented approaches. Established codes such as PeTar \cite{wang_petar_2020}, NBODY6++GPU \cite{wang_nbody6gpu_2015}, and GADGET-4 \cite{springel_simulating_2021} provide state-of-the-art accuracy and performance for large-scale stellar dynamics, but they lack native support for automatic differentiation and gradient calculations. On the other hand, recent differentiable semi-analytic models of stellar streams \citep[][]{alvey_albatross_2024, nibauer_textttstreamsculptor_2024}, demonstrate the power of gradient-based inference, but they necessarily approximate the underlying dynamics. \textsc{Odisseo} bridges these paradigms by implementing a fully differentiable direct N-body solver in JAX, enabling gradients of the final particle state with respect to initial conditions and potential parameters, while also inheriting JAX’s just-in-time compilation and native GPU acceleration. At present, \textsc{Odisseo} uses only direct pairwise force computations with softening, without a tree or fast-multipole algorithm. This ensures exact forces (up to the softening scale), which is advantageous for accuracy and gradient consistency, but comes at the cost of computational scalability, making large-$N$ ($>10^6$) simulations impractical.

\section{Odisseo}

\textsc{Odisseo} is designed to enable inference-driven studies of dynamical systems. A key design choice in \textsc{Odisseo} is to rely on the Python programming language, in contrast to established N-body codes described in Sec \ref{sec:N-body_codes}, which are implemented in low-level languages like C or C++. While these mature codes achieve exceptional performance, their complexity and low-level implementations often make rapid prototyping, unit testing, and community contributions more difficult. By building \textsc{Odisseo} entirely in Python, we prioritize maintainability, readability, and accessibility for a community developments. This choice facilitates faster iteration on scientific ideas, easier integration with modern ML workflows, and a lower barrier to entry for community-driven extensions. Written in a purely functional style and built entirely on the JAX ecosystem, \textsc{Odisseo} provides end-to-end differentiability: the final particle state remains differentiable with respect to all simulation parameters, including initial conditions, integration time, particle masses, and external potential parameters. This is achieved by leveraging JAX’s program transformation capabilities, such as tracing and reverse-mode automatic differentiation, which propagate gradients through the entire simulation pipeline. The functional and modular design, inspired by \cite{storcks_differentiable_2024}, allows components—such as integrators, external potentials, or initial condition generators—to be easily swapped or extended, facilitating rapid experimentation. \textsc{Odisseo} also natively supports just-in-time compilation, automatic vectorization, and execution across CPUs, GPUs, and TPUs, ensuring both flexibility and high performance. Looking forward, \textsc{Odisseo} aims to serve as a community-driven platform for differentiable simulations, adaptable to astrophysical problems such as stellar stream modeling, as well as to other domains where dynamical systems and external potentials play a central role.

\subsection{Distributed performance}\label{sec:benchmark}
\enlargethispage{\baselineskip}
\begin{wrapfigure}[14]{R}{0.40\textwidth}
\vspace{-30 pt}
\textbf{}\includegraphics[width=0.39\textwidth]{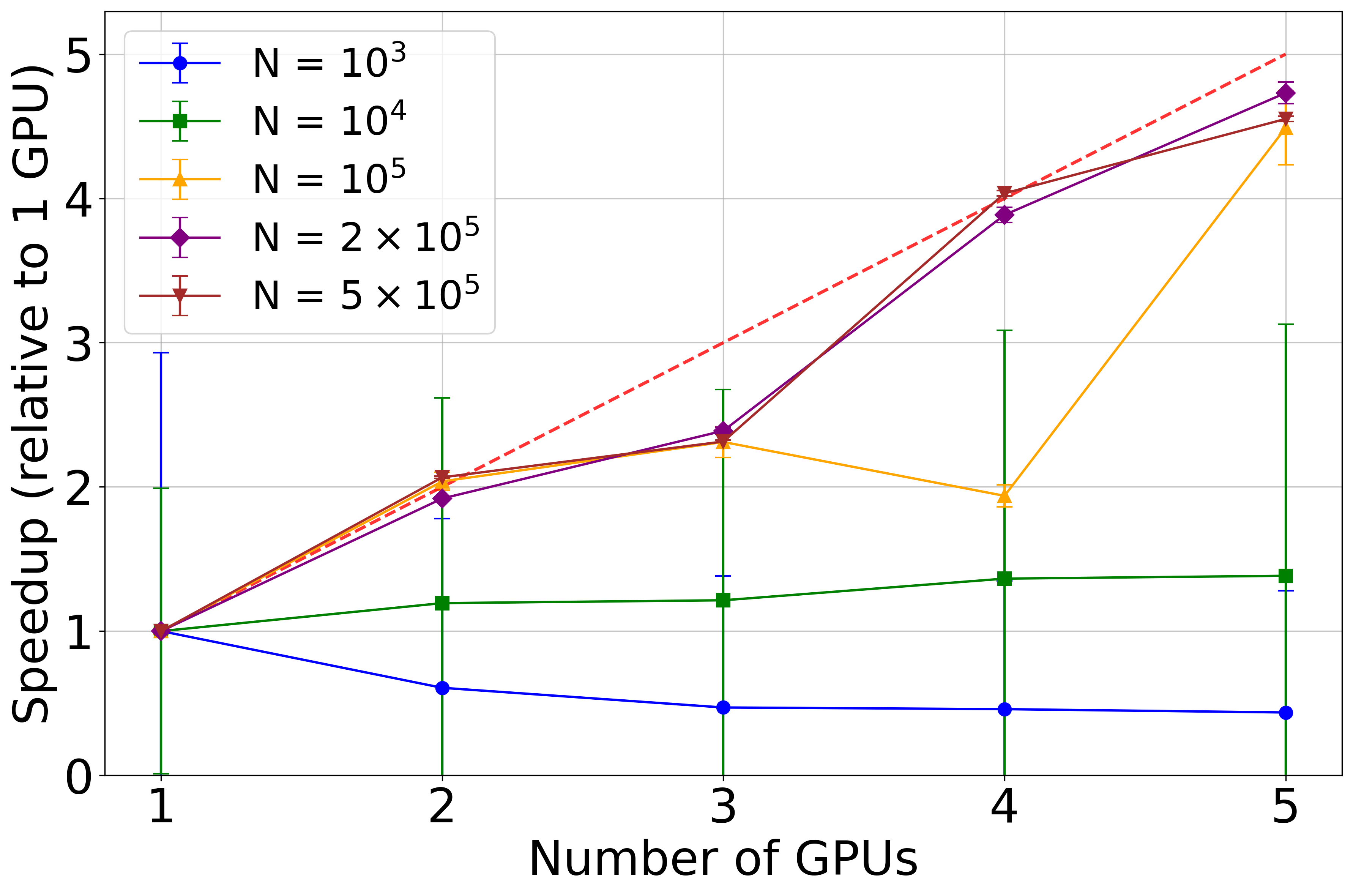}
    \caption{Speedups for a single simulation distributed across multiple GPUs with different number of particles N. The dashed red line is the ideal linear scale, which is achieved for N > $10^4$.}
    \label{fig:scaling}
\end{wrapfigure}

To demonstrate that \textsc{Odisseo} is ready for large-scale parallelization with minimal user effort, we benchmarked it on a standard stellar-dynamical test: the tidal disruption of a Plummer sphere in a realistic Milky Way potential. Leveraging the JAX ecosystem, which provides built-in primitives for distributed data and computation, we distributed a single simulation run across multiple GPUs (NVIDIA A100) without modifying the simulation code. Specifically, we ran the same disruption experiment with increasing particle numbers while distributing the computation over 1 (baseline), 2, 3, 4, and 5 GPUs to measure scaling performance. The results are shown in Fig. \ref{fig:scaling}. Each configuration was repeated three times to report both the mean runtime and its standard deviation. This setup highlights the ease with which \textsc{Odisseo} achieves near-linear strong scaling on modern GPU clusters, confirming its suitability for massively parallel inference-driven simulations.
\begin{figure}
\vspace{-6 em}
\centering
  \includegraphics[width=0.75\textwidth]{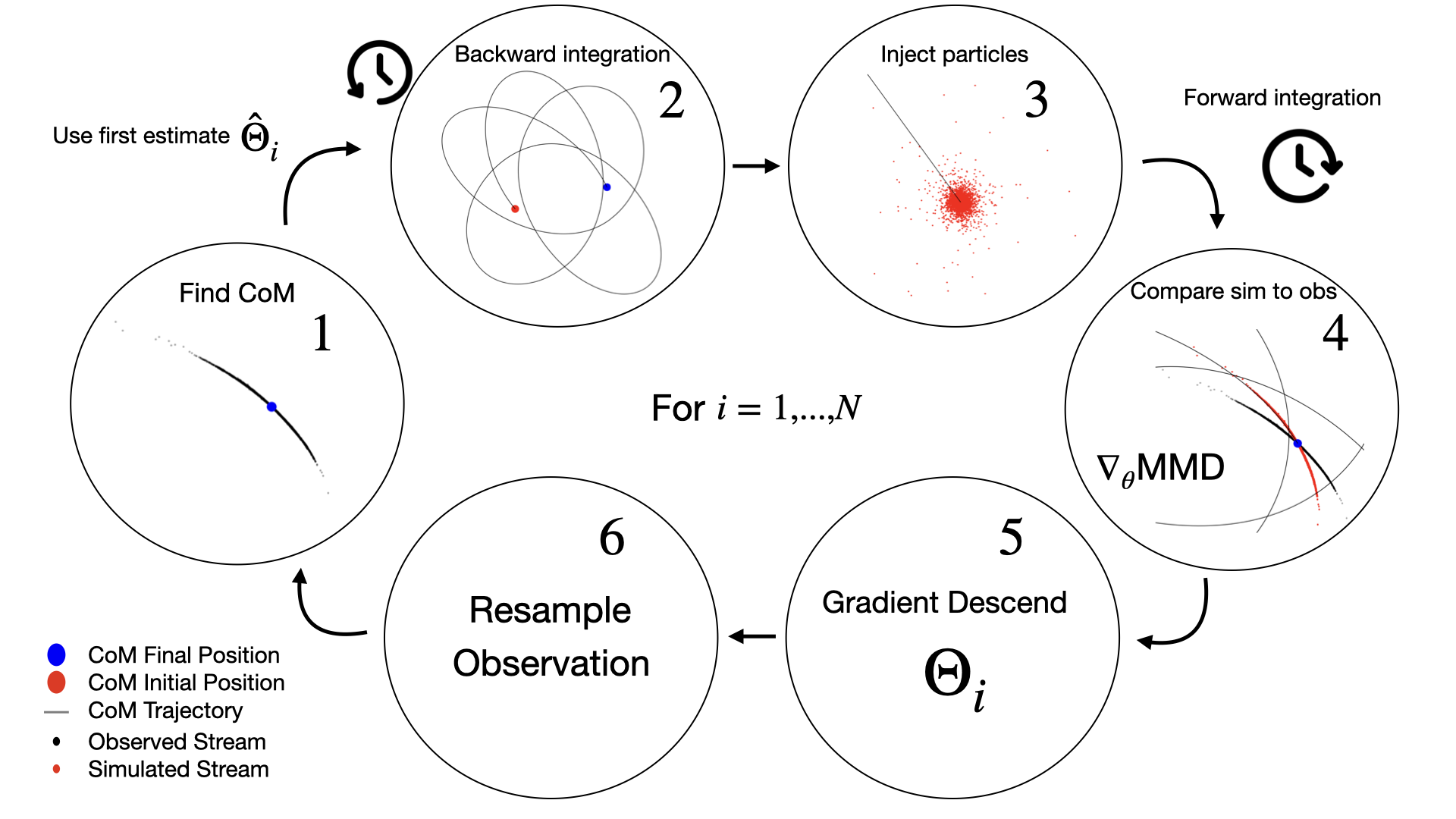}
  \centering
  \caption{Inference pipeline. 
The goal is to generate $N$ bootstrap samples of the parameters. 
1. From the observed stellar stream, we identify its Center of Mass (CoM). 
2. Using an initial estimate $\hat{\Theta}_i = (\hat{t}_{\text{end}}, \hat{M}_{\text{Plummer}}, \hat{M}_{\text{NFW}}, \hat{M}_{\text{MN}})$, obtained through a grid search, we integrate the present-day CoM position backward in time as a single particle of mass $\hat{M}_{\text{Plummer}}$ for a duration $\hat{t}_{\text{end}}$, within an external potential parametrized by $\hat{M}_{\text{NFW}}, \hat{M}_{\text{MN}}$. 
3. The recovered initial CoM position is populated with 1000 particles sampled from a Plummer sphere in equilibrium. 
4. The resulting simulated stream is compared to the observed stream using the Maximum Mean Discrepancy (MMD), and the gradient of the loss is computed. 
5. We perform 10 steps of vanilla Gradient Descent, repeating steps 2--4 internally. The parameters at the minimum MMD are taken as the final estimate for the sample $\Theta_i = (t_{\text{end}}, M_{\text{Plummer}}, M_{\text{NFW}}, M_{\text{MN}})$. 
6. The observed stream is resampled according to observational errors to generate a new realization of the target, and the process is repeated from step 1.}
\vspace{-3.5 em}
\label{fig:flow_chart}
\end{figure}

\subsection{Inference}\label{sec:Inference}
We demonstrate two applications of gradient-based optimization for studying stellar streams with a differentiable N-body code. Importantly, we do not work with real GD-1 observations, but with a mock simulated stream for which the true parameters are known.
The first method, described in Sec.~\ref{sec:gradient_descent_bootstrapping}, introduces a novel approach that treats both observational and simulated data as point clouds and minimizes their discrepancy directly.
The second method, inspired by previous work on stellar streams \citep{koposov_constraining_2010}, assumes that stream stars follow orbits similar to that of their progenitor. This allows the inverse problem to be reformulated as a least-squares minimization between the observed stars and the simulated orbit.

\subsubsection{Gradient Descent and Bootstrapping}\label{sec:gradient_descent_bootstrapping}

\begin{wrapfigure}[26]{R}{0.55\textwidth}
    \centering
    \vspace{-15pt}
    \includegraphics[width=0.54\textwidth]{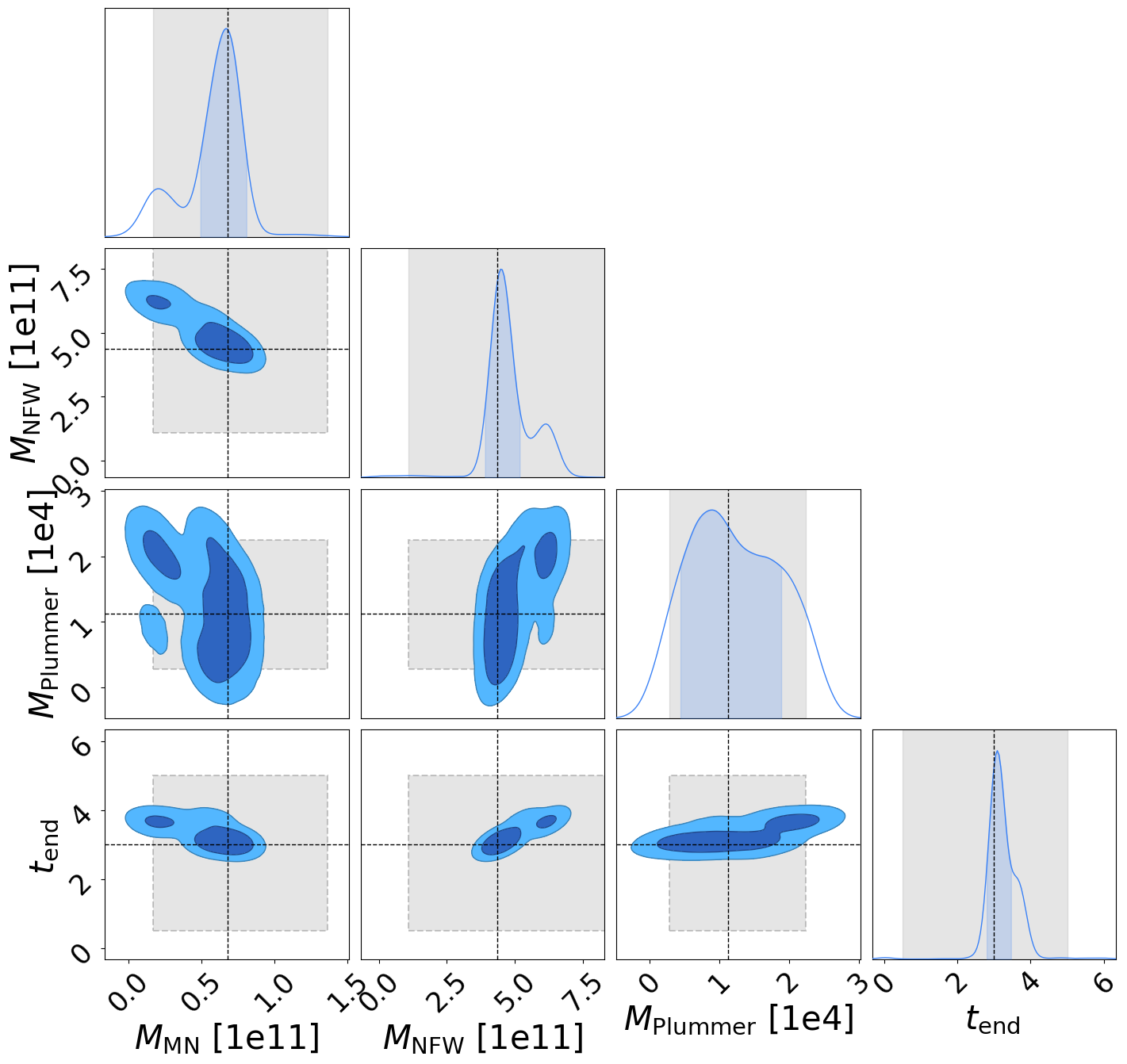}
    \caption{Bootstrapping inference results. The masses $M_{\text{MN}}$, $M_{\text{NFW}}$ and $M_{\text{Plummer}}$ are in $M_\odot$, while the total time of integration $t_{\text{end}}$ is in $Gyr$. The gray box represents the "prior" used to set up the initial grid search. The darker area in the diagonal plots represents one standard deviation from the maximum likelihood. The dashed lines indicate the true values.}
    \label{fig:corner_plot}
    \vspace{-20pt}
\end{wrapfigure}

To illustrate a simple inference example using our differentiable N-body simulator, we use the full 6D phase-space information of particles to recover the parameters $\Theta$ that best describe the host galaxy potential—specifically, the masses of the NFW halo and the MN disk—as well as the progenitor properties, including the Plummer sphere mass and the total integration time, which are respectively $\Theta = (t_{\text{end}}, M_{\text{Plummer}}, M_{\text{NFW}}, M_{\text{MN}})$ .
Following the general setup of \citep{alvey_albatross_2024}, we adopt the \texttt{MWPotential2014} model from \texttt{galpy} \citep{bovy_galpy_2015} as the analytic Milky Way potential.

The overall inference pipeline is summarized in Fig.~\ref{fig:flow_chart}.
Starting from the present-day position and velocity of the stream’s progenitor, we integrate its center of mass backward in time as a single particle to obtain the initial conditions for the simulation.
Unlike the semi-analytic model of \citep{alvey_albatross_2024}, the progenitor is then replaced by 1000 particles drawn from an equilibrium Plummer sphere, shifted in phase space by the backward-integrated initial conditions.
The system is evolved forward in time using a second-order Velocity–Verlet integrator with 1000 fixed time steps.
A few representative snapshots and the progenitor orbit are shown in Fig.~\ref{fig:GD1}.
For the inference, we employ a bootstrapping procedure: the simulated GD-1 stream used as “observation” is resampled 1000 times according to the observational uncertainties reported in \citep{alvey_albatross_2024}, and the inference is repeated for each realization.
The pipeline consists of two stages: first, a grid search over $10^4$ parameter combinations using the Maximum Mean Discrepancy (MMD) between simulated and observed streams as the loss function; second, 10 refinement steps of gradient-based optimization with a learning rate of $10^{-5}$ to improve upon the grid-search results.
Finally, the distribution obtained by bootstrapping is summarized by fitting a kernel density estimate (KDE) to the inferred parameters, from which we generate the corner plot in Fig. \ref{fig:corner_plot}. The experiment was run on 2 NVIDIA A100. The code to reproduce the results is available on GitHub: \url{https://github.com/vepe99/Odisseo/blob/main/notebooks/dev/albastross/uncertainty_quantification.py}


\subsubsection{Orbit fitting and Fisher contours}

\begin{wrapfigure}[25]{R}{0.57\textwidth}
    \centering
    \vspace{-5 pt}  
    \includegraphics[width=0.56\textwidth]{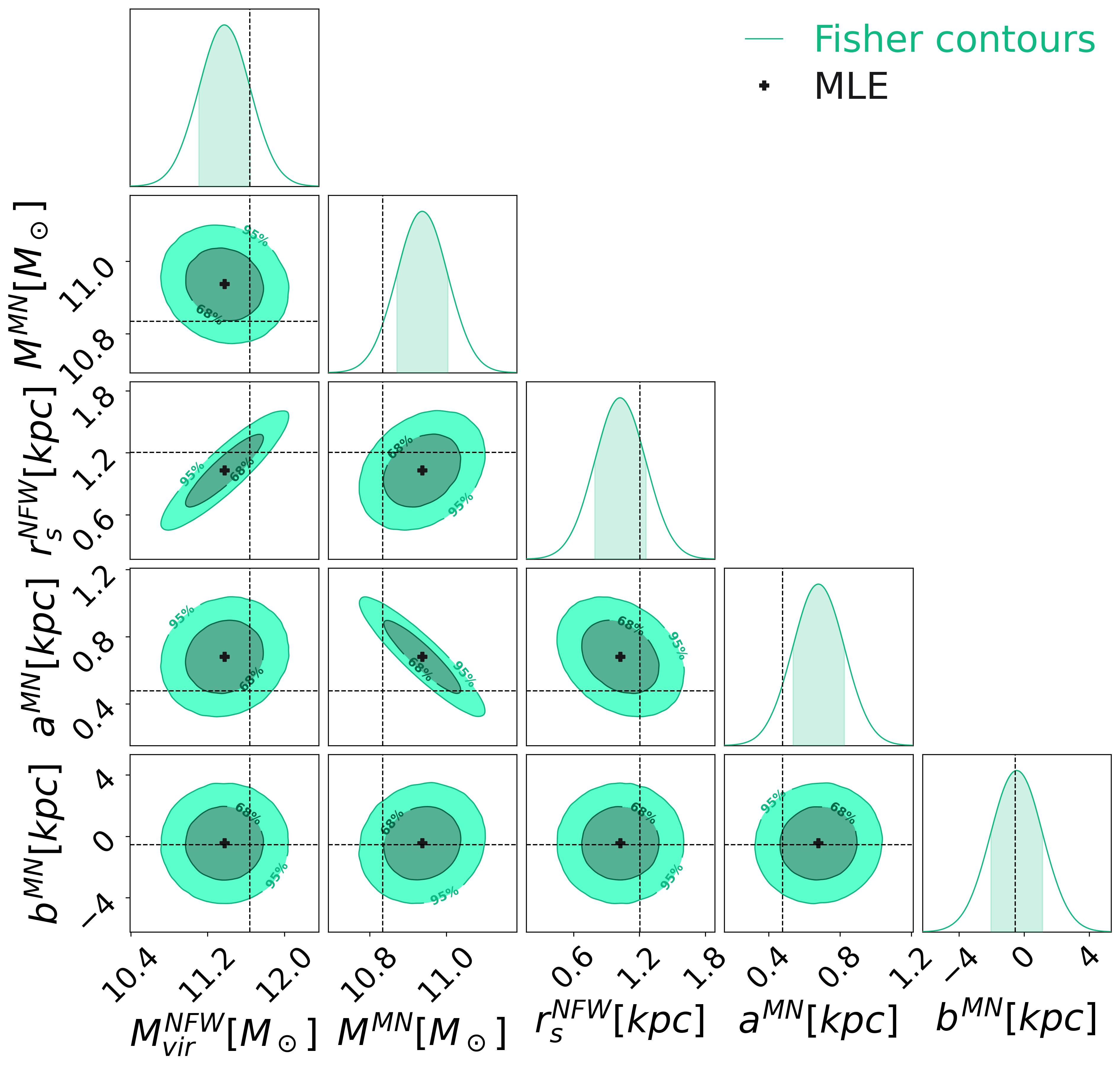}
    \caption{Fisher confidence contours around the Maximum Likelihood Estimate (MLE). All parameter values are shown in logarithmic (dex) units. The dashed lines are indicating the position of the true value.}
    \label{fig:Fisher_contour}
\end{wrapfigure}

For this method, we evolve only a single particle--the progenitor-- with mass $M_{\mathrm{Plummer}} = 10^{4.5}\, \mathrm{M}_\odot$ and compare its orbit with the observed stream, following the setup described in \citep{Mestre_2024}. The integration is  both forward and backward in time for $\Delta t = 2\,\mathrm{Gyr}$, starting from its present-day position.\footnote{For the GD-1 stream, this assumption does not strictly hold since the progenitor is fully dissolved. However, in principle, the initial phase-space coordinates could be included as additional free parameters to infer.} This integration yields an orbit $(\mathbf{X}(t), \mathbf{V}(t))$, which we project onto the stream’s reference frame defined by $(\phi_1, \phi_2, \tilde{\mu}_{\phi_{1}}, \mu_{\phi_{2}}, v_h)$, introduced by \citep{koposov_constraining_2010}. In this frame, $\phi_1$ and $\phi_2$, corresponding respectively to the longitude and latitude in the GD-1 celestial coordinate system, are expressed in radians, $\tilde{\mu}_{\phi_{1}} = \mu_{\phi_{1}}\cos\phi_{2},$ and $\mu_{\phi_{2}},$ are proper motions in $\mathrm{mas\,yr^{-1}}$, and $v_h$ is the heliocentric line-of-sight velocity in $\mathrm{km\,s^{-1}}$. Following the same observational window and data treatment as in \citep{Mestre_2024}, we interpolate the modeled orbit along $\phi_1$ to match the sampling of the observed stream. The goodness of fit is then quantified through a chi-squared statistic, and adopting the same error values as \citep{Mestre_2024}. The resulting $\chi^2$ serves as the log-likelihood, $\mathcal{L} \propto -\chi^2$, for the parameter inference. To obtain the Maximum Likelihood Estimate (MLE), we apply the Levenberg--Marquardt least-squares method implemented in \texttt{optimistix} \citep{optimistix2024} from 1000 randomly sampled initial positions $\theta_i$ (with $0.5 \, \theta_{True}<\theta_i < 2\theta_{True}$) and then take the value with the lowest $\chi^2$. The confidence intervals are derived from the inverse of the Fisher information matrix, assuming a Gaussian approximation of the likelihood around the MLE with fixed covariance matrix. The Fisher matrix is defined as $F_{ij} = \frac{1}{2} \frac{\partial^2 \chi^2}{\partial \theta_i \, \partial \theta_j}$, where $\theta_i$ and $\theta_j$ denote the model parameters, in our case the NFW virial mass $M_{vir}^{NFW}$ and scale radius $r_{s}^{NFW}$, the MN mass $M^{MN}$, scale height $b^{MN}$ and length $a^{MN}$. The Fisher matrix is trivially obtained using  \texttt{jax.hessian}. The covariance matrix of the estimated parameters is then given by $C = F^{-1}$. The results are reported in Fig. \ref{fig:Fisher_contour}. The code to reproduce the results is available on GitHub: \url{https://github.com/vepe99/Odisseo/blob/main/notebooks/dev/albastross/Fisher_contour/Least_square/least_square_search_paper.py}.

\section{Conclusion and outlook}\label{sec:Conclusion}
We have presented \textsc{Odisseo}, a novel, modular, highly parallelizable, differentiable direct N-body code written in JAX, and demonstrated that even complex tasks—such as retrieving the parameters of the host galaxy and the progenitor of a stellar stream—can be tackled with surprisingly high accuracy. Given the computationally demanding nature of the grid search used in our first example, this inference pipeline should be considered primarily as a proof of concept rather than a fully deployable application. The second approach is strongly limited by the Gaussian approximation and by the non-negligible dependency of the MLE algorithm on the initial guess $\theta_i$. A complete analysis would require a proper treatment of uncertainties, for instance via Hamiltonian Monte Carlo or simulation-based inference with simulator feedback, as suggested in \cite{holzschuh_flow_2024}. Also as future works, we aim to leverage the gradient in order to couple the \textsc{Odisseo} with a Neural Network in order to improve the accuracy of the solver, just like as been shown in \cite{um_solver---loop_2021}.

\section*{Broader impact statement}
The authors are not aware of any immediate ethical or societal implications of this work. This work purely aims to aid scientific research and proposes to apply novel differentiable simulator to unravel the formation history of the Milky Way, its morphology and the family of stellar streams that populate its halo.

\begin{ack}
This work is funded by the Carl-Zeiss-Stiftung through the NEXUS program. This work was supported by the Deutsche Forschungsgemeinschaft (DFG, German Research Foundation) under Germany’s Excellence Strategy EXC 2181/1 - 390900948 (the Heidelberg STRUCTURES Excellence Cluster). We acknowledge the usage of the AI-clusters Tom and Jerry funded by the Field of Focus 2 of Heidelberg University.
\end{ack}

{
\small
\bibliographystyle{plain}
\bibliography{main}
}

\newpage

\appendix

\section{Simulation Output}
\begin{figure}[h]
  \centering
  \includegraphics[width=0.8\textwidth]{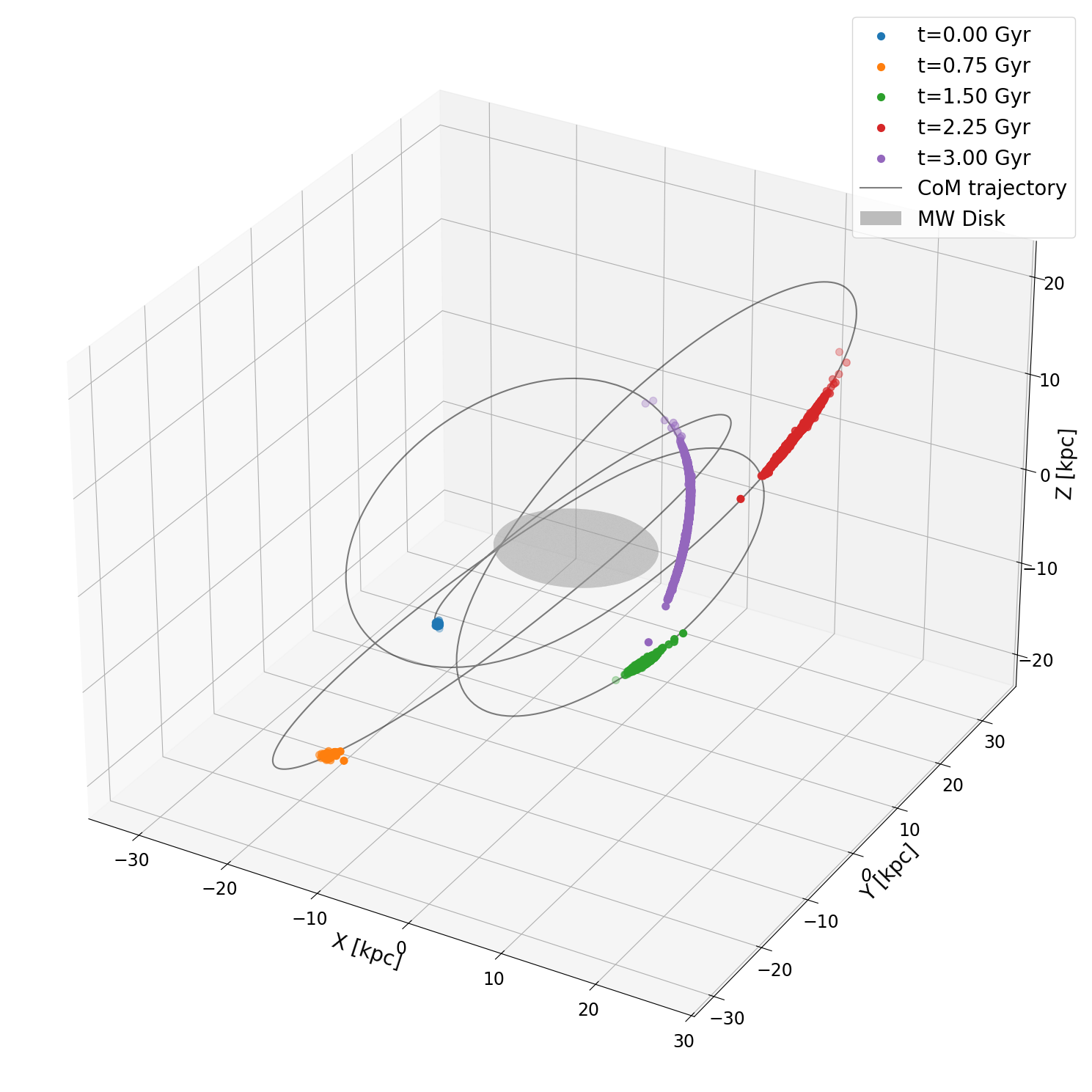}
  \vspace{-1em}
  \caption{Snapshots at different times $t$ of the GD1 stream simulation. Milky Way's stellar disk is shown for scale and the center of  mass (CoM) trajectory is shown as solid line. }
  \label{fig:GD1}
\end{figure}

In this Fig. \ref{fig:GD1} we show a few time steps for the fiducial simulation of the GD1 stream.

\end{document}